\documentclass[aps,prd,floatfix,10pt,nofootinbib,showpacs,notitlepage,twocolumn,superscriptaddress]{revtex4-1}
\usepackage[utf8]{inputenc}

\usepackage{amsmath}
\usepackage{amsfonts}
\usepackage[colorlinks=true,linkcolor=blue,citecolor=blue,urlcolor=blue]{hyperref}
\usepackage{amssymb}
\usepackage{graphicx}
\usepackage{enumerate}
\usepackage{csquotes}
\usepackage{tabularx}
\usepackage{bm}
\usepackage{commath}

\begin{document}

\title[Stress--Impedance Model]{On the use of a Stress--Impedance Model to describe sound propagation in a lined duct with grazing flow}
\author{Yves Aur\'egan}
\email{yves.auregan@univ-lemans.fr}
\affiliation{Laboratoire d'Acoustique de l'Universit\'e du Mans, Centre National de la Recherche Scientifique (CNRS), Le Mans Universit\'e, Avenue Olivier Messiaen, 72085 Le Mans Cedex 9, France}


\date{\today} 

\begin{abstract}
With flow, the acoustical effect of a lined wall cannot be described by a single quantity like the wall impedance. At least two quantities are required. In addition to the impedance, the unsteady tangential force exerted by the wall on the flow has to be taken into account. This force is due either to viscous effects or to the unsteady transfer of axial momentum from the flow into the lined wall. The paper  describes a Stress--Impedance model where the two variables used are the impedance and the friction factor that links the pressure to a tangential stress at the wall. The use of a wall stress helps to better understand the mechanisms of momentum transfer between the flow and the wall in the vicinity of an acoustic treatment.
\end{abstract}


\maketitle

\section{\label{sec:1} Introduction}

Despite its practical importance, the behavior of acoustic treatments in the presence of a grazing flow is still poorly understood. This is due to the complexity of the unsteady turbulent flow near the perforated plate holes that has been demonstrated by numerical simulations \cite{zhang2012numerical}. Most of the currently used models assume that the effect of the flow boundary layer can be described by the Ingard-Myers relation \citep{Ingard1959, Myers1980} and that the flow complexity can be captured by an equivalent impedance that must be empirically or semi-empirically determined. 

Much work has been done to improve the description of the boundary layer effect \citep{khamis2016acoustic}. Despite these advances, the commonly used models are still unable to explain the difference between the impedances deduced from measurements in the flow direction and in the opposite direction \citep{renou2011failure, boden2017comparison, weng2017impedance}.

It has been shown in \citep{rebel1992effect} that the oscillating shear stress can play an important role in sound propagation with a grazing flow along a liner. This shear stress is apparently due to the interaction between turbulent flow and the rough wall which is the interface of the acoustic treatment. Further attempts were made to account for shear stress in terms of viscous stress \citep{Auregan2001, khamis2017acoustics} or in term of additional force acting on the walls of a cavity \cite{kop2008aeroacoustic}. A modified boundary condition was derived that introduces a coefficient $\beta_v$ that characterizes the transfer, by the  normal  fluctuating  displacement,  of  axial  momentum from the steady flow into the lined wall \citep{Auregan2001}.

In this paper, a heuristic approach is used: The existence of a tangential surface force on the wall is postulated and this paper describes how this force can be deduced from measurements or calculations. The detailed analysis of how this shear stress is created at the wall of the liner is outside the scope of this paper and is yet to be investigated. 

The paper is organized as follows: Section~\ref{sec:2} presents the Stress--Impedance model and a way of computing both the impedance and the stress, under the form of an equivalent friction factor, from the knowledge of two different wave numbers in a  two-dimensional (2D) geometry.
Section~\ref{sec:3} presents the application of this model to the numerical simulations made in a 2D propagation problem \citep{dai2016acoustic}.

\section{\label{sec:2} The Stress--Impedance Model}
\subsection{\label{subsec:2:1} General equations}

\begin{figure}[ht]
\includegraphics[width=0.8\columnwidth]{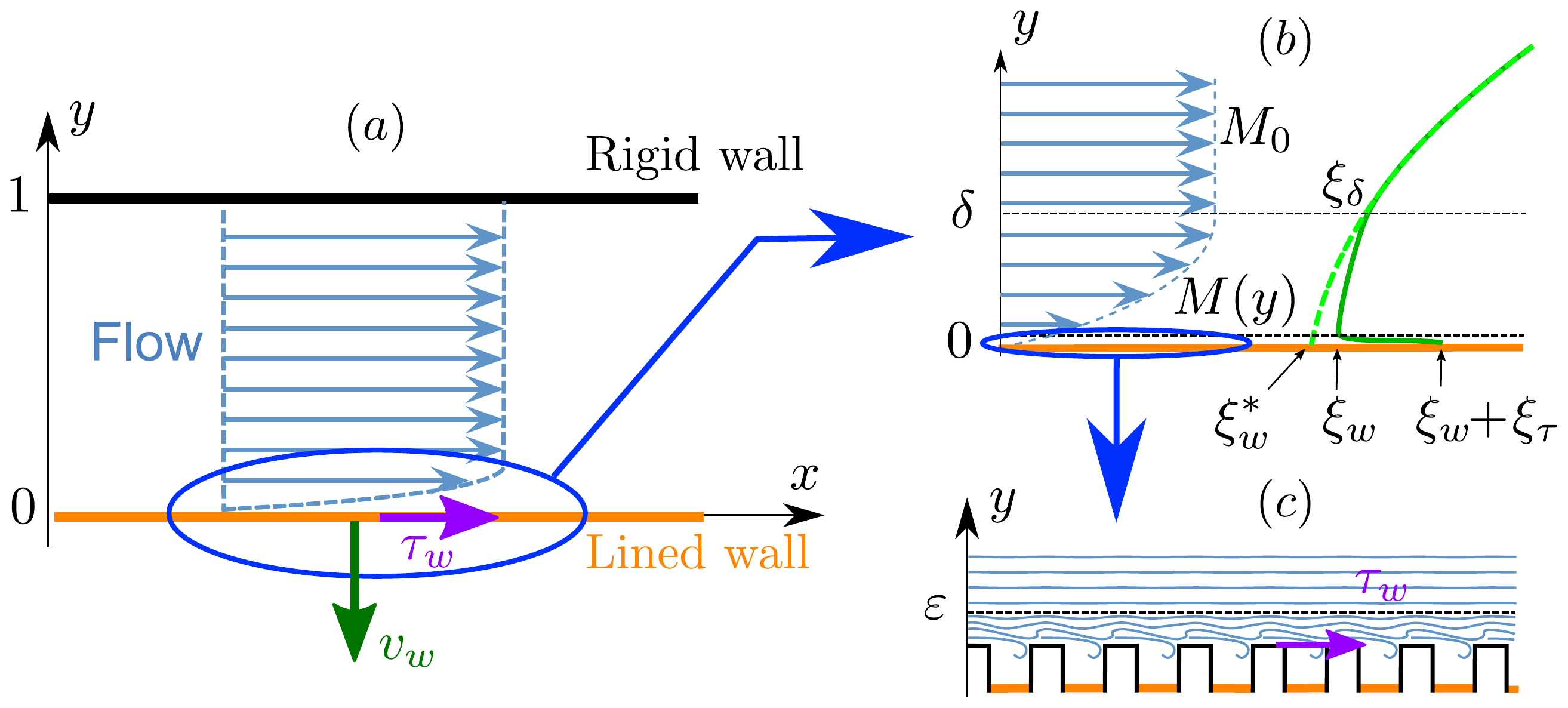}
\caption{\label{fig:FIG1}{(color online) General view of the  2D problem.}}
\end{figure}

The propagation in a two--dimensional (2D) duct of height $h$ with a shear flow of velocity $U(y)$ is considered. The velocity is supposed to be uniform outside of the boundary layer having a thickness $\delta$ and it decreases and then vanishes on the lower wall. On this wall, the duct is acoustically treated. The acoustic treatment is described classically by an admittance $Y_w = v_w/p_w$ that links the normal velocity into the wall $v_w$ to the pressure at the wall $p_w$ but also by a tangential stress $\tau_w$ that is intended to describe an unsteady transfer of momentum from the flow into the wall due to wall roughness and to turbulent and viscous effects. Those effects are supposed to be confined near the wall in a layer of thickness $\varepsilon$  smaller than the mean flow boundary layer $\delta$, see Fig.~\ref{fig:FIG1}. 
To simplify the notations, all parameters are nondimensionalized. All velocities are nondimensionalized by the speed of sound $c_0$, so that the mean velocity becomes the Mach number $M(y)$. Distances are nondimensionalized by the height of the channel $h$, time by $h/c_0$, and pressure by $\rho_0 c_0^2$  where $\rho_0$ is the mean density. 
Except very near the wall ($0<y<\varepsilon$), all the dissipative effects can be neglected and the dimensionless equations governing the acoustic motion are 
\begin{eqnarray}
D_t u + M' v  &=& - \partial_x p \label{eq:1}\\
D_t v &=& - \partial_y p \label{eq:2}\\
D_t p &=& - \partial_x u -\partial_y v \label{eq:3}
\end{eqnarray}
where $u,v$ are respectively the velocities in the $x$ and $y$ direction, $p$ is the pressure,  $D_t=\partial_t+M \partial_x$ is the convective derivative and $M'=d_y M$. 
To avoid singular terms in the above equations when the boundary layer thickness vanishes, it is advantageous to rewrite Eqs. (\ref{eq:1}--\ref{eq:3}) only in terms of pressure $p$ and transverse displacement $\xi$ ($v=D_t \xi$) which are the regular variables in the boundary layer (those variables remain continuous when the  boundary layer thickness vanishes):
\begin{eqnarray}
\partial_y p &=& - D^2_t \xi  \label{eq:4}\\
D^2_t \partial_y \xi &=& \partial^2_x p -D^2_t p \label{eq:5}
\end{eqnarray}
The pressure and the displacement are  taken under the form $p(x,y,t)= \hat{p}(y) \exp(\mathrm{j}(\omega t - k x))$ and $\xi(x,y,t)= \hat{\xi}(y) \exp(\mathrm{j}(\omega t - k x))$ and  Eqs. (\ref{eq:4}--\ref{eq:5}) become: 
\begin{eqnarray}
\mathrm{d}_y \hat{p} &=& \Omega^2 \hat{\xi}  \label{eq:6}\\
\Omega^2  \mathrm{d}_y \hat{\xi} &=& k^2 \hat{p} -\Omega^2 \hat{p} \label{eq:7}
\end{eqnarray}
where $\Omega = \omega - k M(y)$. In the following, the hats are removed for simplicity.

\subsection{\label{subsec:2:2} Effect of the boundary layer}

At the lowest order, when the boundary layer thickness is very small compared to the height of the channel ($\delta\ll 1$), Eqs. (\ref{eq:6}--\ref{eq:7}) show that  $p_w=p^*_w$ and $\xi_w=\xi^*_w$ where $p^*_w$ and $\xi^*_w$ are the values at the wall when the flow is uniform up to the wall. At this level of approximation, the boundary layer only induces a jump in the normal velocity $\omega v^*_w = \Omega_0  v_w$ where $\Omega_0 = \omega - k M_0$ and $M_0$ is the Mach number in the uniform flow. 

A more precise description, at the first order in $\delta$, is obtained by integrating Eqs. (\ref{eq:6}--\ref{eq:7}) over the boundary layer  \citep{brambley2011well}: 
\begin{eqnarray}
p_w-p^*_w &=& \delta \mathcal{I}_0 \Omega_0^2 \: \xi_w  \label{eq:8}\\
\xi_w-\xi^*_w  &=& \delta \mathcal{I}_1 k^2/\Omega_0^2   \: p_w  \label{eq:9}
\end{eqnarray}
where 
$$
\delta \mathcal{I}_1 = \int_0^\delta 1-  \left(\frac{\Omega _0 }{\Omega}\right)^2 \mathrm{d}y
\:\:\: \mathrm{and} \:\:\:
\delta \mathcal{I}_0 = \int_0^\delta  1- \left(\frac{\Omega}{\Omega_0}\right)^2 \mathrm{d}y 
$$

\subsection{\label{subsec:2:3} Stress along the wall}

To study the near wall zone ($0<y<\varepsilon$), the effect of a shear stress along the $x$ direction is added to Eq. (\ref{eq:7}):
\begin{equation}
\Omega^2  \mathrm{d}_y {\xi} = k^2 {p} -\Omega^2 {p} - \mathrm{j} k \mathrm{d}_y \tau \label{eq:11}
\end{equation}
By considering that the mean flow is very weak in the near wall zone and that $\varepsilon\ll 1$, this equation can be integrated along $y$ in $\omega^2 \xi_\tau = \mathrm{j} k \tau_w$ where $\tau_w$ is the stress at the wall and $\xi_\tau$ is an additional displacement due to the stress.
It can be noted that this relation has to be modified ($\omega$ become $\omega - U_s k/c_0$ ) if a slip velocity $U_s$ is considered at the wall to take into account the effect of the roughness on the turbulent motion \cite{schlich}. 

\subsection{\label{subsec:2:4} Equivalent boundary condition}

At the lowest order, when the thickness  of the boundary layer $\delta$ is negligible, the relation between $p^*_w$ and $v^*_w$, pressure and normal velocity at the wall when a perfect uniform flow is considered, and $p_w$ and $v_w$, pressure and normal velocity at the wall when the boundary layer and the stress are considered, are  
\begin{equation}
p^*_w = p_w 
\:\:\: \mathrm{and} \:\:\:
 v^*_w = \frac{\Omega_0}{\omega} v_w - \frac{\Omega_0k}{\omega^2} \tau_w
\end{equation}

In this case, the equivalent admittance $Y_w^*=-v_w^*/p_w^*$ (seen by a wave propagating in a uniform flow) can be computed from the admittance of the wall ($Y_w=-v_w/p_w$) and $f_w$ by
\begin{equation}
Y^*_w = \frac{\Omega_0}{\omega} \left(Y_w + \frac{k}{\omega} f_w \right) \label{eq:12}
\end{equation}
where $f_w = \tau_w/p_w$ can be seen as an equivalent friction coefficient.

In the uniform flow,  Eqs. (\ref{eq:6}--\ref{eq:7}) result in $\mathrm{d}^2_y p = -\alpha^2 p$ where $\alpha^2 = \Omega_0^2 -k^2$. The pressure can be written $p=A \cos(\alpha (1-y))$ and, at the lined wall $y=0$, the relation between pressure and velocity is $-v(0)/p(0)=Y^*_w =  -\mathrm{j} \alpha \tan(\alpha)/ \Omega_0$.

When two values of the wavenumber $k$ are known, two values of $Y^*_w$ can be computed and Eq. (\ref{eq:12}) can be used for the determination of the admittance $Y_w $ and of the friction coefficient $f_w$. For a more precise solution, the influence of the boundary layer can be taken into account by \citep{brambley2011well}:

\begin{equation}
\left( 1 + \frac{\mathrm{j}  \Omega_0^2}{\omega }  \delta \mathcal{I}_0 Y_w \right)Y^*_w = 
\frac{\Omega_0}{\omega} \left(Y_w + \frac{k}{\omega} f_w \right) + \mathrm{j} \frac{k^2}{ \Omega_0} \delta \mathcal{I}_1\label{eq:13}
\end{equation}

\subsection{\label{subsec:2:5} Links with previous formulations}

When the Ingard-Myers boundary condition is used without taking into account the surface force, Eq.~(\ref{eq:12}) applies with $f_w = 0$.  In some papers \cite{Auregan2001, renou2011failure}, the Ingard-Myers  condition has been modified by using a new parameter $\beta_v$ and the relation between the normal velocity at the wall with an uniform flow $v^*_w$ and the normal velocity in the liner $v_w$ is transformed into $v^*_w=(1-(1-\beta_v) M_0 k/\omega) v_w $. Eq.~(\ref{eq:12}) becomes:
\begin{equation}
Y^*_w = \frac{\Omega_0}{\omega} Y_\beta + \frac{k M_0} {\omega}\beta_v Y_\beta  \label{eq:Eq7}
\end{equation}

In general, the admittance $Y_w$ extracted using the Stress--Impedance model is not equal to the admittance $Y_\beta$ extracted using the $\beta_v$ model.
Thus the two formulations are not exactly equivalent and will give different results.  

\section{\label{sec:3} Application of the Stress--Impedance model}
\subsection{\label{subsec:3:1} Numerical determination of the wavenumbers}
\begin{figure}[ht]
\includegraphics[width=\columnwidth]{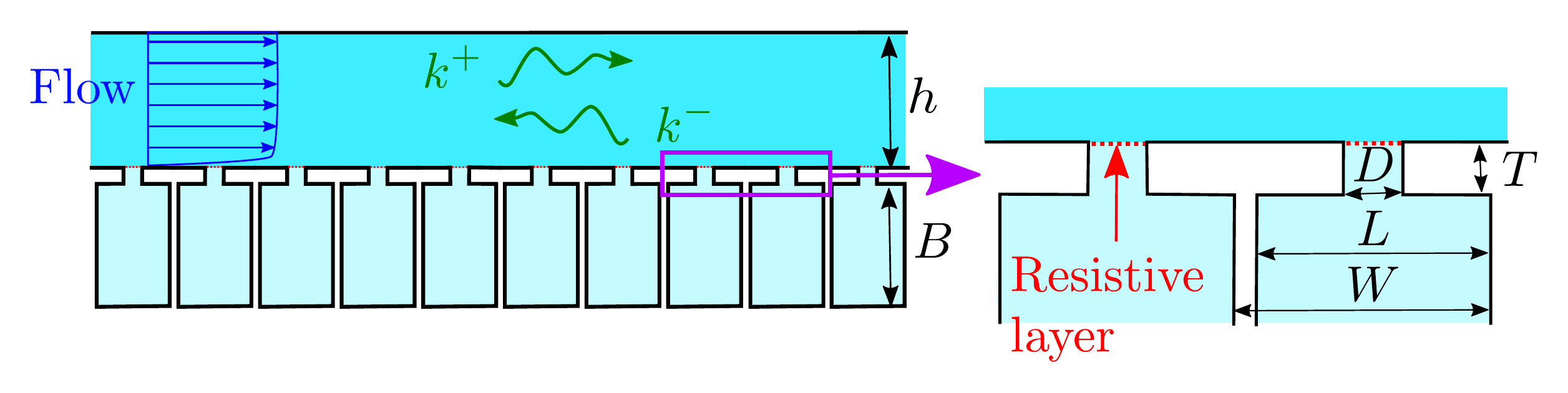}
\caption{\label{fig:FIG2}{(color online) Sketch of an array of 2D periodic cells with Helmholtz resonators.}}
\end{figure}

The linear acoustic propagation with flow is computed in an array of 2D periodic cells with Helmholtz resonators, see Fig.~\ref{fig:FIG2}. Neglecting the viscous and thermal losses, the linearized Euler equations are solved in one of the periodic cells by using the multimodal method as described in \citep{dai2016acoustic}. In the numerical calculation, the geometry is defined by the height of the duct $h$ = 15 mm, the depth of the resonator cavity $B$ = 25 mm, the period between two resonators which is equal to the width of the cavity $W=L$ = 5 mm, the thickness of the resonator neck $T$ = 0.5 mm and the width of the hole $D$ = 1 mm. 
With those dimensions, the resonance frequency of the Helmholtz resonators is 2700 Hz.
A resistive layer has been added to insure some dissipation in the neck of the Helmholtz resonators and the normalized resistance is 0.05. A shear flow profile has been taken into account and the Mach number is given as a function of the mean Mach number $M_0$ by $M(y) = M_0 (m+1)(1-(1-y)^m)/m$ where  $m=30$ to insure a small thickness of the boundary layer. 

\begin{figure}[ht]
\includegraphics[width=\columnwidth]{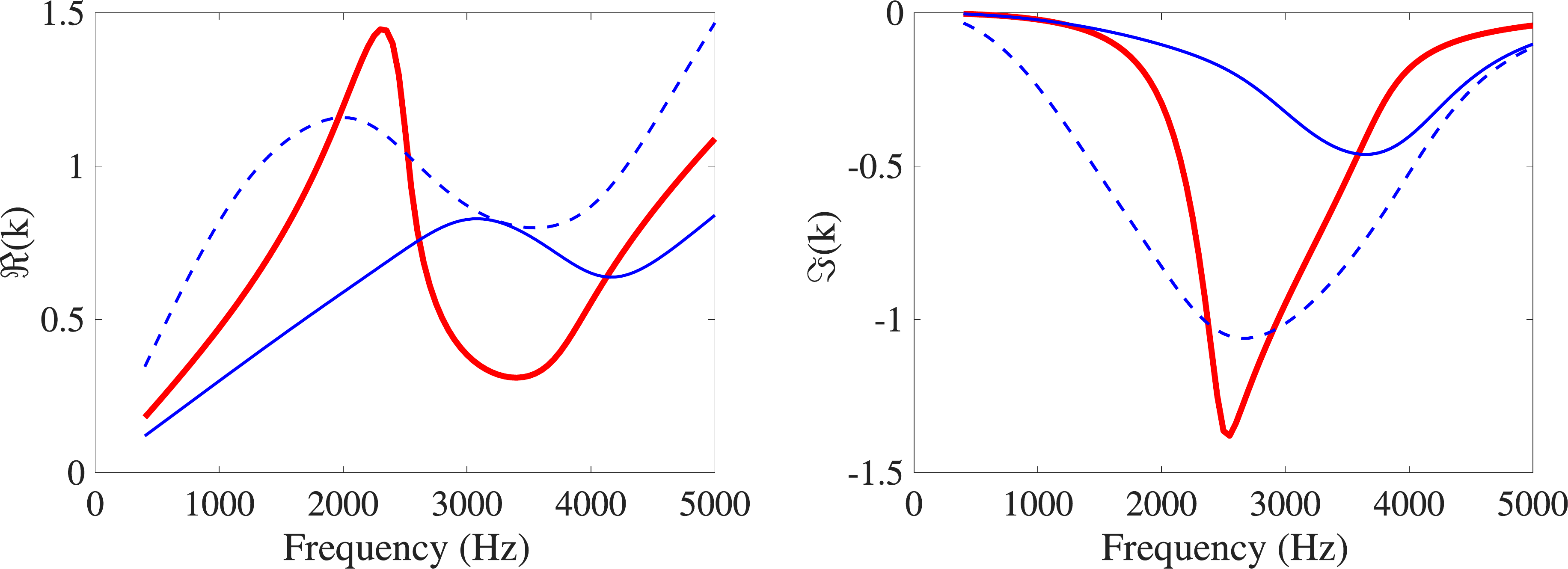}
\caption{\label{fig:FIG3}{(color online) Real and imaginary values of the dimensionless wavenumbers $k_B^+ $ (continuous lines) and $-k_B^- $ (dashed lines) without flow (thick red curve) and with flow ($M_0=0.3$, in blue). Without flow $k_B^+=-k_B^-$.}}
\end{figure}

The output of the numerical calculations is a transmission matrix that links all the modes at the entrance of one cell to the modes at the exit of this cell. In the present calculation, 900 modes been considered. Using the Floquet-Bloch approach, the wavenumbers in the periodic system are computed and the wavenumbers of the least attenuated modes in each propagation direction $k_B^+$ and $k_B^-$ are selected and will be used in the following to compute the impedance and the friction coefficient $f_w$.  The value of $k_B^+$ and $k_B^-$ are plotted in Fig~\ref{fig:FIG3}.

\subsection{\label{subsec:3:2} Impedance and friction factor}

\begin{figure}[ht]
\includegraphics[width=\columnwidth]{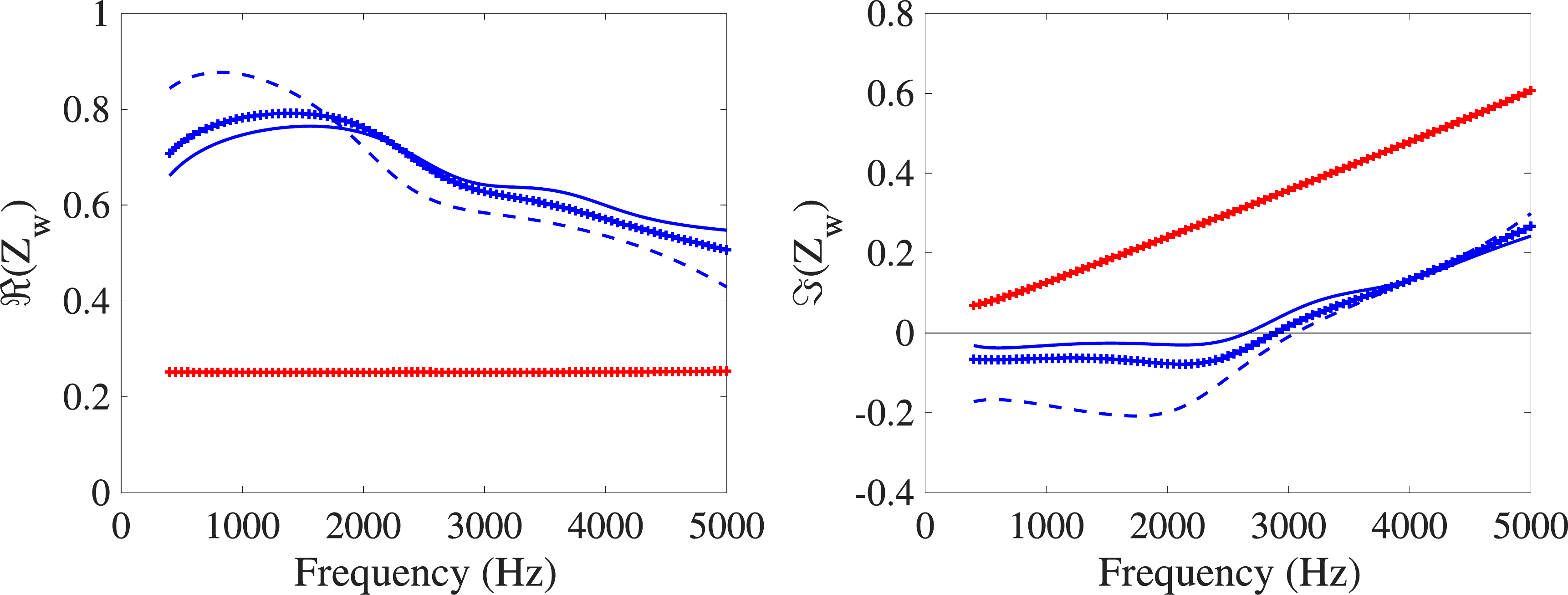}
\caption{\label{fig:FIG4}{(color online) Real and imaginary values of the equivalent impedance without flow (in red) and with flow ($M_0=0.3$, in blue). The symbols represents the value computed with the Stress--Impedance model, the continuous line (resp. dashed line) is the value obtained by considering $f_w=0$ from $k_B^+$ (resp. $k_B^-$).}}
\end{figure}

When the boundary layer effect is neglected, Eq.~(\ref{eq:12}) is written, using $k=k_B^+$ and $k=k_B^-$, 
\begin{equation}
Y^\pm= Y_w+ k_B^\pm f_w/\omega  \label{eq:Eq8} 
\end{equation}
where $Y^\pm= -\mathrm{j} \omega \alpha^\pm \tan(\alpha^\pm)/ (\Omega_0^\pm)^2$. The two Eqs.~(\ref{eq:Eq8}) allow the determination of $ Y_w$ and $f_w$. The value of the dimensionless impedance of the plate is computed by removing the effect of the cavity from the impedance of the resonator: $Z_w=1/Y_w+\mathrm{j}/\tan(\omega B)$. It is plotted in Fig~\ref{fig:FIG4} and compared to the two values $Z^\pm$  obtained from $Y^\pm$  by assuming $f_w=0$ in Eq.~(\ref{eq:Eq8}). 
Without flow, the three values of the impedance are equal. The real part is almost constant and equal to the resistance of the dissipative layer divided by the percentage of open area. The imaginary part increases linearly with the frequency and is related to the mass of fluid moving in the hole and its vicinity. With flow, the two impedances $Z^\pm$ deduced by assuming $f_w=0$ are different showing again that the equivalent impedance depends on the direction of the incident waves in the classical approach \citep{dai2016acoustic}. On the contrary, the impedance is determined  in a unique way in the Stress--Impedance model.

\begin{figure}[ht]
\includegraphics[width=\columnwidth]{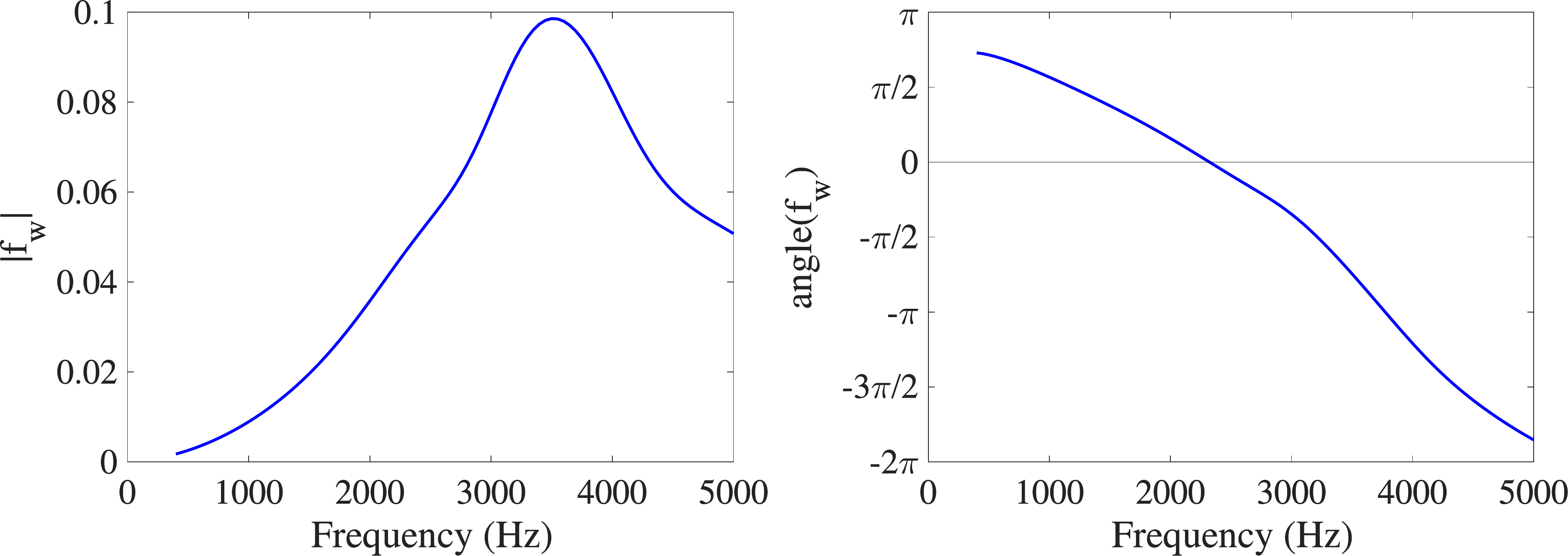}
\caption{\label{fig:FIG5}{(color online) Modulus and phase of the friction coefficient $f_w$ in the case with flow ($M_0=0.3$).}}
\end{figure}

The additional effect, which is supposed to describe the difference between impedances with different wave incidences (i. e. different values of $k$), is the tangential force acting on the lined wall. It is described by the friction coefficient $f_w$ which is plotted in Fig~\ref{fig:FIG5}.

\begin{figure}[ht]
\includegraphics[width=0.7\columnwidth]{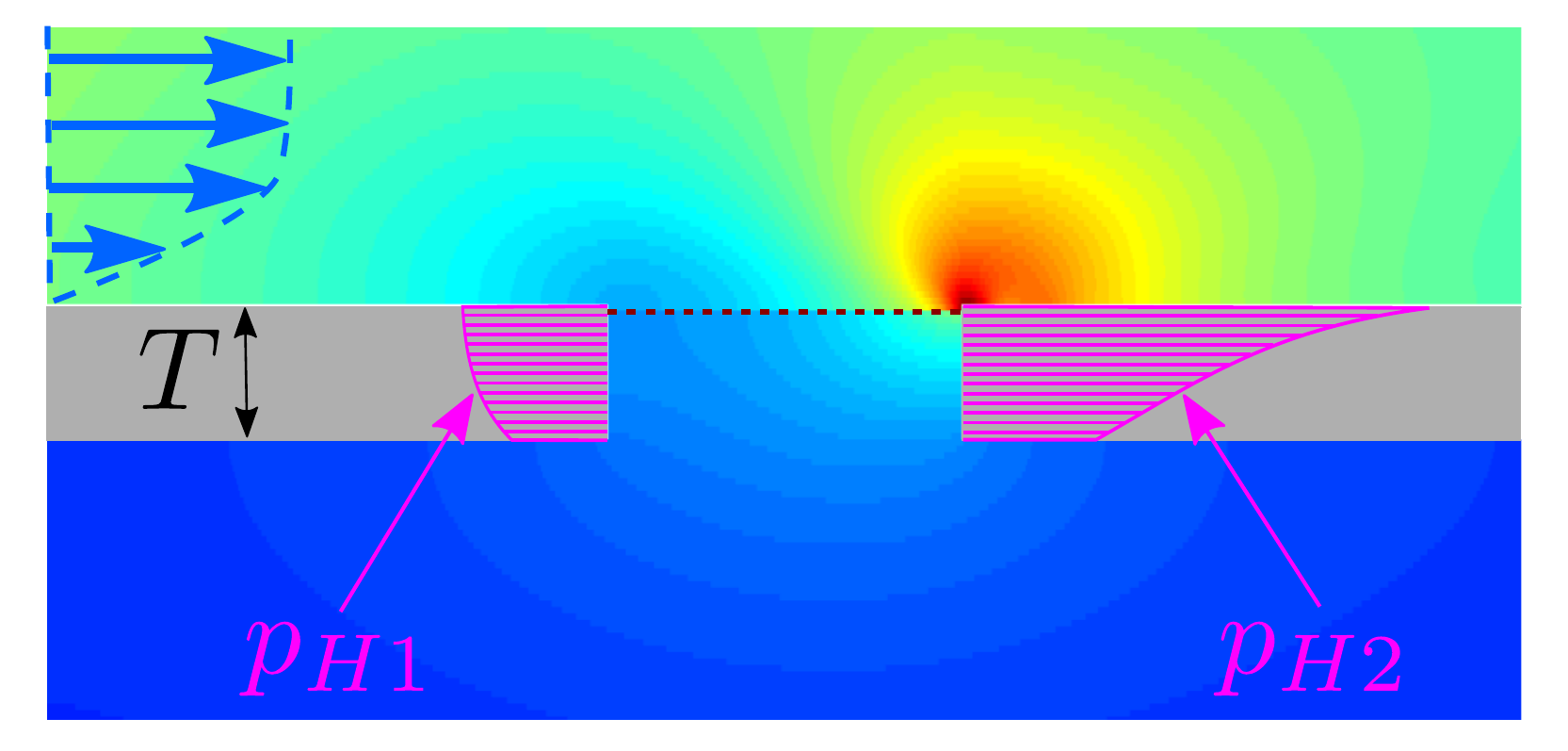}
\caption{\label{fig:FIG6}{(color online) Pressure field computed in a hole at 3600 Hz. The magenta curves ($p_{H1}$ and $p_{H2}$) represents the pressure distribution on the vertical walls of the hole.  }}
\end{figure}

The amplitude of the friction coefficient  starts from 0 and increases to reach a maximum amplitude at 3.5 kHZ, which is slightly higher than the resonance frequency of Helmholtz resonators with flow (3.3 kHz). The phase of the friction coefficient indicates that stress and pressure are in opposition of phase at low frequencies. The phase decreases regularly and the opposition of phase occurs again at 3.8 kHz. This indicates that there is a characteristic time delay (0.26 ms) between stress and pressure. Looking at Fig~\ref{fig:FIG6}, it can be thought that a part of this stress comes from the unsteady force applied to the vertical walls of holes by an hydrodynamic mode that is created at the level of the upstream wall and that is convected and amplified.  This convection time may explain the delay between stress and pressure. 
However, the results of the numerical simulation should be interpreted with caution because an artificial damping of the hydrodynamic modes has been added near the rigid wall to mimic the destruction of coherent structures by turbulence (see Fig. 3 in \citep{dai2016acoustic}). This damping results in an artificial change in momentum along the $x$ direction. A more precise numerical simulations (possibly including turbulence, viscous and thermal effects) will have to be carried out to analyze more precisely the forces exerted by the lined wall on the fluid.

The acoustical effect of a lined wall with grazing flow cannot be described by a single quantity like the wall impedance. At least two quantities are needed. The two quantities used in the Stress--Impedance model are the impedance and a friction factor which links the pressure to a tangential stress at the wall. Compared to previous two quantities model like  \citep{Auregan2001}, the use of a wall stress can help to better understand the mechanisms of momentum transfer between the flow and wall in the vicinity of an acoustic treatment.

\begin{acknowledgments}
This work was supported by the "Agence Nationale de la Recherche" international project FlowMatAc No. ANR-15-CE22-0016-01.
The author benefits from fruitful discussions with J. Golliard, G. Gabard and X. Dai.
\end{acknowledgments}

\end{document}